\documentclass[twocolumn,showpacs,showkeys,preprintnumbers,amsmath,amssymb]{revtex4}

\usepackage{setspace}
\usepackage{graphicx}% Include figure files
\usepackage{dcolumn}% Align table columns on decimal point
\usepackage{bm}% bold math
\usepackage{epsfig}
\begin{document}

%\doublespacing

\preprint{Internal Report submitted to Review$/DF-IST$}

\title[]{Electromagnetotoroid Structures and their Hydrodynamic Analogs}

\author{Mario J. Pinheiro} \email{mpinheiro@ist.utl.pt}
\affiliation{Department of Physics Instituto Superior T\'{e}cnico, Av. Rovisco Pais,
1049-001 Lisboa, Portugal \\
\& Institute for Advanced Studies in the Space, Propulsion and Energy Sciences
265 Ita Ann Ln.
Madison, AL 35757
USA
}

\thanks{We gratefully acknowledge partial financial support by the
Reitoria da Universidade T\'{e}cnica de Lisboa and Funda\c{c}\~{a}o para a Ci\^{e}ncia e Tecnologia (FCT).}


%\subjclass{}


\date{\today}
%\dedicatory{}%
%\commby{}
% ----------------------------------------------------------------
\begin{abstract}
We introduce the concept of an electromagnetotoroid in astrophysics, and explore its role in polar jets. This model represents the onset of Abraham's force driven by some external source, for example, the infall of gas towards a star. The Abraham's force term is analogous to the Magnus force, and thus represents the formation of electromagnetic vortex structures in the fabric of space-time. In principle, the proposed toroidal field structure can also provide force spaceship propulsion.
\end{abstract}

\maketitle
% ----------------------------------------------------------------

\section{Introduction}

Kronenberg {\it et al.}~\cite{Kronberg_01} suggested that the magnetic field lines of galaxies extend a few million light years into the intergalactic medium. Although the mechanism it is not still fully understood, black hole accretion disk energy could be converted into magnetic fields through an efficient dynamo (within charged black holes), a kind of cosmic electric motor. Occasionally, accretion disk systems eject huge amounts of gas. For example, certain Active Galactic Nucleus expel jets of plasma into space, a phenomena first observed by Alan Marscher's team from the object {\it BL Lacertae}. The plasma jet from this system spirals outward from a flattened disk of spinning gas surrounding a supermassive black hole, and extends 950 million light years beyond the host galaxy~\cite{Marscher}.

Newly formed stars  (``pre-T-Tauri")  are usually surrounded by bipolar jets and molecular outflows in regions with small patches of nebulosity. Such stars are known as Herbig-Haro (HH) objects. Several models have been proposed to explain their jet ejection-accretion processes, and it is becoming evident that pure hydrodynamical models are not sufficient. According to MHD simulations, magneto-centrifugal ejection may be the driving mechanism~\cite{Cabrit}. Optical observations~\cite{Hartigan_2004} indicate that jets are produced in regions 5.5 AU in diameter, while attaining distances of 800 AU from the source. This geometry is a typical Mach angle for free lateral expansion of a supersonic jet~\cite{Cabrit}.

Returning to Earth, atmospheric phenomena include two broad classes of lightning-like flashes: sprites~\cite{Sentman_93} and elves~\cite{Lyons_94}. These short-lived, luminous structures are associated with the convective cells of large thunderstorms.
Brief flashes of light in the stratosphere above thunderstorms were first predicted by C. T. R. Wilson~\cite{Wilson_25}. The smallest sprites, named C sprites, are probably single vertical columns. They can gather together with downward-branching tendrils (called jellyfish), or exhibit upward branching toward the ionosphere. They apparently share a similar fundamental mechanism. In fact, Watanabe~\cite{Watanabe_99} has presented optical data supporting the conclusion that ``column-sprites" are always preceded by elves.

The results cited above illustrate the view that electric currents pervade the universe, and, that a mechanism exists capable of expelling  matter to astronomical distances. Aside from the fact that electromagnetic fields are involved, however, the mechanism of jet accretion in Herbig-Haro stars and active galactic nuclei remains a mystery.

In biology, several creatures use a propulsion mechanism that relies on the production of vortices by wings, paddles and fins~\cite{Dickinson}. Fish swim by flapping their tail and other fins, creating vortices in the water that carry away momentum. Squid and salps move by ejecting fluid intermittently, producing vortex rings, with the shape that gives the maximum thrust for a given energy input~\cite{Turner}. Quite interestingly, all these creatures shape their path through the water while the vortex produced at each stroke go behind them, like a motion sustained by traveling through a channel of vortices.

In this article, we intend to give evidence that the vortex creation mechanism can not only explain the propulsion used by living beings on Earth, but also the jets created by cosmic events. The electromagnetic field, side by side with the fluidic Magnus force, broadens our view of the problem. In particular, we will derive and discuss the nature of the electromagnetotoroid vortex structure, aiming to develop previous work on fluidic electrodynamics~\cite{Pinheiro2,AlexPin1}.

\section{The electromagnetotoroid structure}

It has been shown that in the natural world, propulsion through a fluid medium relies on the production of vortices by a material structure~\cite{Dickinson} (e.g., wings, paddles, fins). We intend to show that this general mechanism, can also account  for the electrodynamic acceleration of fluids, by reaction against the physical vacuum~\cite{HarpazSoker,Feigel04,Pinheiro1}, a plasma, or any other kind of fluid~\cite{Pinheiro2}.

It is known from electrodynamics, that the ponderomotive force acting on the material of an electromagnetic propelling device is provided by Abraham's force density, $\mathbf{f}^A$ (e.g., Refs.~\cite{ShockleyJames,Feigel04,Pinheiro1,Pinheiro2}).

\begin{figure}
  \includegraphics[width=3.5 in]{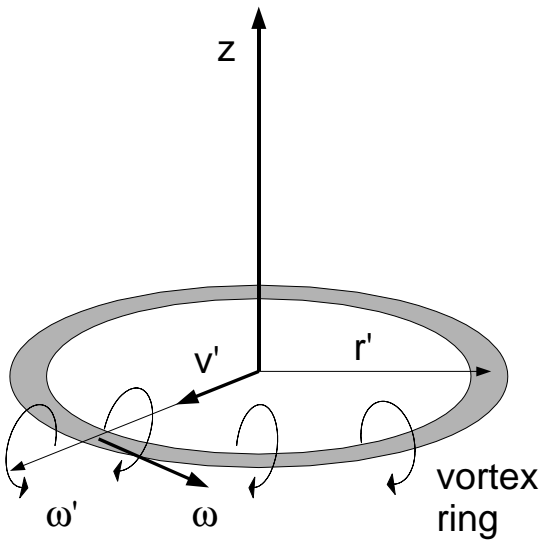}\\
  \caption{Vortex ring whose radius varies with the radial velocity $v_r$.}\label{fig1}
\end{figure}

It is also known that, in the framework of hydrodynamics, the three-dimensional Magnus force is given by~\cite{Huggins}:
\begin{equation}\label{eq0}
    \mathbf{f}_{Mxy} = -\rho [(\mathbf{V}_v -\mathbf{V}_{vo}) \times \mathbf{k}_z],
\end{equation}
where $\mathbf{V}_v$ is the velocity of the vortex center of mass, $\mathbf{V}_{vo}$ is the stream velocity, $\rho$ is the fluid density, and $\mathbf{k}_z$ is a vector oriented along the $z$ direction with magnitude equal to the circulation. In this paper, we show that Magnus and Abraham forces represent the same type of structure in the fabric of space-time: a vortex capable of propelling material bodies.

First, note that the ``magnetic current force", produced by the magnetic
charges that ``flow" when a magnetic field changes, is given by
$\mathbf{f}_m=\varepsilon_0 [\mathbf{E} \times (\mathbf{\dot{B}} -
\mu_0 \mathbf{\dot{H}})]$\cite{ShockleyJames}. This is the ``Abraham term" appearing in the Abraham force density $\mathbf{f}_A$, which differs
from the Minkowsky force density $\mathbf{f}_M$ by means of the
expression (see also Refs.~\cite{Feigel04,Brito}):
\begin{equation}\label{eq1}
\mathbf{f}_A=\mathbf{f}_M + \frac{\varepsilon_r \mu_r
-1}{c^2}\frac{\partial }{\partial t} [\mathbf{E} \times \mathbf{H}].
\end{equation}
The last term of Eq.~\ref{eq1} corresponds to the {\it vacuum-interactance}~\cite{Clevelance}, which is associated with the momentum as follows:
\begin{equation}\label{eq2}
\mathbf{g}^V=\frac{\epsilon_r \mu_r -1}{c^2}[\mathbf{E} \times
\mathbf{H}].
\end{equation}
We know that a magnetic dipole at rest $\mathfrak{M}$ in an external
(and homogeneous) electric field $\mathbf{E}$
has momentum given by $[\mathfrak{M} \times \mathbf{E}]/c^2$. When the magnetism of the dipole changes, the density of force is given by the last term of
Eq.~\ref{eq1}. The Abraham term represents the force transmitted to a material structure (see also Ref.~\cite{Pinheiro 2011}).

The analogue hydrodynamic force analogous to Abraham's force is the
Magnus force, given by Eq.~\ref{eq0}. In the next lines, we aim to show how the Abraham's force can propel a plasma jet.

In the natural world, fishes and birds propel themselves through a liquid medium by using their limbs to transfer momentum to the liquid via vortex structures. Their basic mechanism is to form a vortex structure (e.g.: Ref.~\cite{Hedenstrom}) analogous to that depicted in Fig.~\ref{fig1}. The Magnus force is given by $\rho [\mathbf{V}' \times \pmb{\Gamma}]$ (by unit of length), the force vector being perpendicular to both $\mathbf{V}'$ and to the vortex ``eye".

We will now look at the mechanism that generates the Magnus force. First of all, it is necessary to consume energy in order to progressively enlarge the vortex with a characteristic radial velocity $\mathbf{V}'$ (see Fig.~\ref{fig1}). The toroidal structure is a vortex ring formed by a closed vortex tube of a given diameter (let us say $\chi$). As is well-known in fluid dynamics, this structure is very stable.

The duty mechanism that provides this radial velocity (inward or outward from the central axis) may have different sources. One example is gas falling into stars, in the case of polar jets; Another is the sharp increase of electric current generated by the growing magnetic field of a plasma.

With this kind of mechanism, we can associate a given circulation $\mathbf{\Gamma}'$ (due eventually to an induced field $\mathbf{\omega}'$). The falling (or expelled) stream of particles, most probably will acquire a curved trajectory and   angular momentum, all effects concurring to the formation of the ring with circulation $\mathbf{\Gamma}$ (and vorticity $\mathbf{\omega}$).

At the core of the vortex structure, the resultant force $\mathbf{F}_A$ is aligned along the $Z$ axis. Newton's third law predicts a mecahnical reaction force $\mathbf{F}_{mec}$, which can propel a device (or a magnetized fluid). Therefore, we must have $\mathbf{F}_{mec}=-\mathbf{F}_A$ (in Fig.~\ref{fig1}, $\mathbf{F}_{mec}$ represents the mechanical force pointing downward).

Now, we intend to show that there is an electrodynamical counterpart - Abraham force - which plays an analogous in the formation of an electromagnetoroid. According to our model, such structures can be responsible for jet propulsion by HH objects. It is possible that the same concept could be applied to provide electromagnetic propulsion for a spaceship.

Let us now explore the concept in more detail. Firstly, replace the ``hydrodynamic magnetization" term in Eq.\ref{eq2} with the {\it constitutive relationship} $\mathbf{M}=\eta \mathbf{\omega}$, where $\eta$ represents a given property of the  medium (a dimensionaless constant). This mapping gives us the {\it analogous} hydrodynamic force (by unit of length):
\begin{equation}\label{eqhyem}
d \mathbf{F}^H_m= -\frac{\rho}{c^2} \frac{\partial }{\partial t}[\mathbf{\omega} \times \mathbf{l}] dv.
\end{equation}
Here, $\rho$ is the mass density and $dv$ is the differential volume element. We must understand Eq.~\ref{eq2} as representing the interaction of magnitudes fed by {\it different} energy sources: the circulation $\mathbf{\omega}$ is associated with motion around the vortex-ring, while the Lamb vector $\mathbf{l}=[\pmb{\omega} \times \mathbf{v}]$ is associated with the increasing vortex radius. $\mathbf{\omega}$ spirals about the azimuthal direction, forming a closed circular loop around the main axis.

It is interesting to note that Eq.~\ref{eqhyem} points to the existence of dual forces: one dependent on the fluid angular acceleration (or time-dependent magnetic force); the other dependent on the Lamb-vector time dependency (or time-dependent electric field).

Let us use the cylindrical geometry, shown in Fig.~\ref{fig1}, with $\mathbf{\omega}=\omega_{\theta}\mathbf{u}_{\theta}$ and $\mathbf{l}=l_r \mathbf{u}_r$. The total force resulting from this geometry is given by the following expression:
\begin{equation}\label{eq3}
F^H_m = \frac{\rho}{c^2} \int \int \int \omega_{\theta} \frac{\partial l_r}{\partial t} \mathbf{u}_z dr dz r d \theta.
\end{equation}
We can rearrange terms to obtain
\begin{equation}\label{eq4}
F^H_m = \rho \int \omega_{\theta} \oint_{S'} [\nabla \times \mathbf{\omega}'] dz r d \theta dr,
\end{equation}
where we have used the hydrodynamic form of Amp\`{e}re's equation:
\begin{equation}\label{eq5}
\frac{\partial \mathbf{l}}{\partial t} = - c^2 [\nabla \times \mathbf{\omega}'].
\end{equation}
$\mathbf{\omega}'$ now represents a different (axial) vector (than $\mathbf{\omega}$). In fact, it is the vorticity associated with the increasing Lamb vector. The vorticity vector is oriented along the radial axis. $c$ is a characteristic speed of the medium (see Fig.~\ref{fig1}). Hence:
\begin{equation}\label{eq5}
F^H_m = - \rho \int \omega_{\theta} \oint_{\gamma} (\mathbf{\omega}' \cdot d \mathbf{p}) dr.
\end{equation}
However, we also have $\oint_{\gamma} (\mathbf{\omega'} \cdot d\mathbf{p})=v_r'$. We therefore obtain
\begin{equation}\label{eq6}
F^H_m = - \rho \int \omega_{\theta} v'_r dr = -\rho v_r' \int \omega_{\theta} dr.
\end{equation}
The last integral of Eq.~\ref{eq6} is the circulation $\Gamma_{\theta}=\int \omega_{\theta} dr$ (by unit of length). This result can be recast in the following form:
\begin{equation}\label{eq7}
\mathbf{F}^H_m = \rho [\mathbf{v} \times \mathbf{\Gamma}].
\end{equation}
Eq.~\ref{eq7} shows that Abraham's force is the electromagnetic analogue of Magnus's force in hydrodynamics (by unit of length).

Therefore, if the analogy is valid, we conclude that Abraham's force represents a kind of vortex structure formed in the physical vacuum. The associated reaction force can propel a material structure through space. From this general mechanism, we can envisage a mode of spaceship propulsion based on generating electromagnetic vortices, along with the development of high-current accelerators and thermonuclear devices.

A different but related phenomenon is the Herbig-Haro (HH) class objects observed by Sherburne Wesley Burnham~\cite{Burnham}. HH objects are highly ionized, and their jets are highly collimated. In our viewpoint, the jets may be propelled by a mechanism similar to that presented above. Stars in their first hundred thousand years of existence are often surrounded by an accretion disk or torus~\cite{Novikov}, built-up by gas (or plasma) falling into the strong gravitational field. The accretion disk is formed, most probably, because there is an oblate spheroid attracting particles. When particles fall towards the center, the angular momentum associated with the surrounding material flows outward. A proposed mechanism for this effect is MHD turbulence~\cite{Priest_2000}. Accretion disks are not devoid of magnetic fields~\cite{Priest_2000}, since they constitute a current of ionized particles. The rapid rotation of the inner parts of these disks {\it along with} the inflow of ionized gas creates collimated polar jets of partially ionized plasma perpendicular to the disk, a phenomena also known as polar jets~\cite{Bacciotti}. Unfortunately, the symmetry of the fields produced by the electromagnetoroid (i.e., the accretion disk) does not explain how jets can form along {\it both} polar axes.

For an overview of the different processes driving polar jets see, e.g., Ref.~\cite{Ferreira}.

On Earth, the simplest forms of propulsion are inherent to animals moving on solid ground. They push against the ground, and thus creating reaction forces in the opposite direction. Swimming and flying animals use a complex form of locomotion, because their limbs push against a
fluid. When a fin or wing flaps, it generates a pattern of wake vortices ({\it Von Karman streets}). In general, each stroke forms a discrete vortex similar to a smoke ring~\cite{Dickinson}. The vortex induces a jet flow, which conveys momentum to the fluid. The average force with which an animal propels itself through the fluid is related to the size, strength and velocity of the vortices generated during each stroke~\cite{Hedenstrom,Dickinson}.

We may try to evaluate the advantages of both kind of forces (electromagnetic and hydrodynamic) that may result from the proposed mechanism, using as parameter the ``performance" factor $\eta \equiv mgv/P_{EM,hyd}$. The power associated to the electromagnetic force (as given by Eq.~\ref{eq1}) is $P\approx \epsilon_r \omega \beta E B /\mu_o c$. For a discharge electric current with charge average speed corresponding to a relativistic factor $\beta=v/c=10^{-6}$, and an electric field of the order of $E=10^6$ (V$/$m), assuming that $\epsilon_r=1$ and an external magnetic field of the order of $B=2$ T, the proposed mechanism may deliver the power $P \approx 3$ MW, which may impart to a spacecraft with one tonne of mass the average speed of 300 m$/$s, approaching Mach 1, and $\eta_{EM}=1$. On the other hand, considering Eq.~\ref{eq0}, we have $F \sim \Gamma vR$, and assuming that all dimensions are of the order of unity ($R \sim 1$ m), then, it follows $F \sim \omega^2$, which gives a power output of $P \sim 27$ MW, for $\omega \sim 300$ rad$/$s, assuming the same order of magnitude as before for the speed of the fluid around the airfoil (e.g., ~\cite{27}), and with $\eta_{hyd}=0.11$. We may additionally remark that in a rotating plasma configuration the relative permittivity can attain $\epsilon_r = 10^6$~\cite{Anderson 1959}.
These comparison, while challenging, may ultimately provide the best framework to outline propulsion devices and, particularly, to speculate about advanced propulsion concepts~\cite{Glen}.

\section{Conclusion}

It is generally accepted that the Abraham term represents the force transmitted to a material structure. We have shown in this paper that Abraham's force is the analogue of the Magnus force, and thus represents the formation of vortex structures in the electromagnetic field and physical vacuum. On Earth, vortices transmit momentum and are used by animals to propel themselves through a fluid medium. Therefore, this mechanism is worthy of investigation as a possible major source for astrophysical jets and a potential technique for spaceship propulsion.

\begin{acknowledgments}
The author gratefully acknowledge partial financial support by the
FCT (Funda\c{c}\~{a}o para a Ci\^{e}ncia e a Tecnologia).
\end{acknowledgments}

% ----------------------------------------------------------------
%\INPUT{Xbib.bib}   % For Gather Purpose Only
%\INPUT{Doc2.bbl}  % For Gather Purpose Only
%\bibliographystyle{amsplain}
%\bibliography{Doc2}
\end{document}